# Complete Solution of the XXZ-Model on Finite Rings. Dynamical Structure Factors at Zero Temperature

K Fabricius † and K-H Mütter §

Physics Department, University of Wuppertal
42097 Wuppertal, Germany

U Löw ‡

Institut für Physik, Universität Dortmund
44221 Dortmund, Germany

**Abstract.** The finite size effects of the dynamical structure factors in the XXZ-model are studied in the euclidean time ($\tau$)-representation. Away from the critical momentum $p = \pi$ finite size effects turn out to be small except for the large $\tau$ limit. The large finite size effects at the critical momentum $p = \pi$ signal the emergence of infrared singularities in the spectral ($\omega$)-representation of the dynamical structure factors.

Short title: Dynamical Structure Factors of the XXZ-Model

† Klaus.Fabricius@wptu13.physik.uni-wuppertal.de
§ muetter@wpts0.physik.uni-wuppertal.de
‡ loew@wptu14.physik.uni-wuppertal.de

## 1. Introduction

Based on a complete diagonalization of the XXZ-Hamiltonian:

$$H = 2\sum_{x=1}^{N}[s_1(x)s_1(x+1) + s_2(x)s_2(x+1) + \cos\gamma s_3(x)s_3(x+1)] \qquad (1.1)$$

on finite rings with $N = 4, 6, ..., 16$ sites and anisotropy parameter $\gamma/\pi = 0.0, 0.1, 0.3, 0.4, 0.5$ we reported in References [1] and [2] on the thermodynamics and the static structure factors at finite temperature. In this paper we continue our numerical investigation of the XXZ- model with an analysis of the dynamical structure factors:

$$S_j(\omega, p, T, \gamma, N) = Z^{-1}\sum_n \delta(\omega - (E_n - E_m))\delta(p - p_n + p_m) \times \exp\left(-\frac{E_m}{T}\right)|\langle n|s_j(0)|m\rangle|^2, \quad j = 1, 3. \qquad (1.2)$$

$Z$ is the partition function and $|n\rangle$ denotes an eigenstate of the Hamilton operator and of the momentum operator with eigenvalues $E_n$ and $p_n$, respectively. The dynamical structure factors contain the information on the transition probabilities $|\langle n|s_j(0)|m\rangle|^2$ between the eigenstates $n$ and $m$ with an excitation energy $\omega = E_n - E_m$ and a momentum transfer $p = p_n - p_m$. At $T = 0$ and $p = \pi$ the XXZ-model is known to be critical and here one expects that quantum effects become most important. Therefore, most of the previous studies were concentrated on the grounstate behaviour. There exist analytical results on the dynamical correlation functions for the special case $\gamma = \pi/2$, i.e for the XX-model [3-6]. This model can be mapped on a free fermion system [7] by means of a Jordan Wigner transformation. For the general case $0 \leq \gamma < \pi/2$, the spectrum of the lowlying excitation energies has been exploited by des Cloiseaux and Pearson [8]. In particular it was found from the Bethe ansatz solution that there is a lower bound

$$E_n - E_m \geq E_1(p_1, \gamma) - E_0(p_0, \gamma) = \omega_1(p, \gamma) = \frac{\pi}{\gamma}\sin\gamma\sin p \qquad (1.3)$$

for the excitation energies depending on the momentum transfer $p = p_1 - p_0$. Approaching the boundary (1.3) the structure factors diverge [9]:

$$S_j(\omega, p, T = 0, \gamma, N = \infty) = A\frac{\Theta(\omega - \omega_1)}{(\omega^2 - \omega_1^2)^{\alpha_j(\gamma)}}, \qquad (1.4)$$

where

$$\alpha_1(\gamma) = \frac{1}{2}\left(1 + \frac{\gamma}{\pi}\right), \quad \alpha_3(\gamma) = \frac{\pi/2 - \gamma}{\pi - \gamma}. \qquad (1.5)$$

Starting from these rigorous results G. Müller and collaborators [10] made an ansatz for the dynamical structure factors which has been applied successfully on the description of neutron scattering data [11].

The outline of the paper is as follows. In Section 2 we present our results on the dynamical structure factors at $p = \pi$ and $N = 4, 6, ..., 16$. We make use of the euclidean time representation, which is particularly suited to analyze finite size effects. In Section 3 we demonstrate how infrared singularities emerge on finite systems. In Section 4 we report on our results for noncritical momenta $p < \pi$.

## 2. Finite Size Behavior of the Dynamical Structure Factors at $p = \pi$ and $T = 0$

In the critical regime:

$$p \to \pi, \quad T \to 0, \quad N \to \infty, \tag{2.1}$$

the dynamical structure factors (1.4) develop infrared singularities. We want to investigate how these singularities show up on finite systems, where the structure factors are sums of $\delta$-function contributions. Finite size effects are not so easy to analyze and we therefore look for a smoothening procedure which allows to extract the thermodynamical limit from finite systems. For this purpose let us consider the Laplace transforms of the structure factors. At $T = 0, p = \pi$ they acquire the following form:

$$S_j(\tau, p = \pi, T = 0, \gamma, N) = \sum_{n>0} \delta(p_n - p_0 - \pi) e^{-\tau(E_n - E_0)} |\langle n | s_j(0) | 0 \rangle|^2. \tag{2.2}$$

The variable $\tau$ can be interpreted as an euclidean time. At $\tau = 0$ we recover the static structure factors which behave for $N \to \infty$ as [12]:

$$S_j(\tau = 0, p = \pi, T = 0, \gamma, N \to \infty) = r_j(\gamma) \frac{\eta_j(\gamma)}{\eta_j(\gamma) - 1} \times \left[ 1 - \left( \frac{N}{N_j(\gamma)} \right)^{1 - \eta_j(\gamma)} \right], \tag{2.3}$$

with critical exponents [9]:

$$\eta_1(\gamma) = \eta_3(\gamma)^{-1} = 1 - \frac{\gamma}{\pi}. \tag{2.4}$$

According to (2.3,4) the longitudinal structure factor stays finite whereas the transverse one diverges. This divergence originates from the infrared singularity (1.4) which is integrable for the longitudinal case ($2\alpha_3(\gamma) < 1$) but nonintegrable in the transverse case ($2\alpha_1(\gamma) > 1$). On finite systems the infrared singularities are not visible directly due to the gap:

$$\omega_1(p = \pi, \gamma, N) = E_1 - E_0 = O(N^{-1}) \tag{2.5}$$

between the groundstate energy $E_0$ and the energy $E_1$ of the first excited state.
In the following we consider the Laplace transform (2.2) normalized to the corresponding static structure factor:

$$R_j(\tau,\gamma,N) = \frac{S_j(\tau,p=\pi,T=0,\gamma,N)}{S_j(\tau=0,p=\pi,T=0,\gamma,N)}, \quad j=1,3. \tag{2.6}$$

By construction these ratios are monotonically decreasing with $\tau$ and varying between 1 and 0 for $\tau > 0$. Figures 1(a),(b);2;3(a),(b) present the ratio (2.6) for the longitudinal case at $\gamma/\pi = 0.5, 0.3$, the isotropic case $\gamma = 0$, and the transverse case at $\gamma/\pi = 0.3, 0.5$, respectively. Going from figure 1(a) to figure 3(b) we observe the following characteristic features:

(i) Finite size effects are small in the longitudinal case at $\gamma/\pi = 0.5$ and increase with decreasing values of $\gamma$.

(ii) Finite size effects are large in the transverse case and increase with increasing values of $\gamma$.

(iii) The 'half width' $\tau(R_j = 0.5,\gamma)$- defined as the value of $\tau$, where the ratio (2.6) has dropped to $R_j = 0.5$- moves systematically:

$$\begin{aligned} &\tau(R_3 = 0.5, \gamma = 0.5\pi) < \tau(R_3 = 0.5, \gamma = 0.3\pi) < \tau(R = 0.5, \gamma = 0) \\ &< \tau(R_1 = 0.5, \gamma = 0.3\pi) < \tau(R_1 = 0.5, \gamma = 0.5\pi). \end{aligned} \tag{2.7}$$

As will be shown below, this property is related to the strengthening of the infrared singularity (1.4) according to (1.5):

$$\alpha_3(0.5\pi) < \alpha_3(0.3\pi) < \alpha(0) < \alpha_1(0.3\pi) < \alpha_1(0.5\pi).$$

The integrability of the infrared singularity in the Laplace transform (2.2) for the longitudinal case means that the ratio (2.6):

$$R_3(\tau,\gamma,N\to\infty) = R_3(\tau,\gamma) \tag{2.8}$$

converges to a scaling curve $R_3(\tau,\gamma)$. The large $\tau$-behaviour of the scaling curve is given by the infrared singularity (1.4):

$$R_3(\tau,\gamma) \stackrel{\tau\to\infty}{\longrightarrow} \tau^{2\alpha_3-1}. \tag{2.9}$$

We have tried to determine the inverse $\tau = \tau(R_3,\gamma)$ of the scaling curve from a finite size analysis of our results for $N = 4, 6, ..., 16$ with a parametrization:

$$\ln\tau(R_3,\gamma,N) = \ln\tau(R_3,\gamma) + \frac{B_3(R_3,\gamma)}{N^{\beta_3(R_3,\gamma)}} \tag{2.10}$$

We fixed the parameters $\tau(R_3,\gamma), B(R_3,\gamma), \beta(R_3,\gamma)$ at $N = 12, 14, 16$. We checked the validity of (2.10) by comparison with our low $N$ data ($N = 4, 6, 8, 10$). The dashed curves in figures. 1(a),(b) represent the resulting scaling curve $R_3(\tau,\gamma)$ for $\gamma = 0.5\pi$ and $\gamma = 0.3\pi$, respectively. The exponent $\beta_3(R_3,\gamma)$ decreases with decreasing values of $R_3$ and $\gamma$, which is a signal for increasing finite size effects. This feature is easy to understand. With increasing values of $\tau$ the low energy excitations get a stronger weight in the Laplace transform (2.2). On the other hand, the spectrum of low energy excitations is particularly sensitive to the finiteness of the system. We can compare our scaling curve $R_3(\tau,\gamma)$ with the ansatz of Müller et. al. [10] for the longitudinal structure factors:

$$S_3(\omega, p=\pi, T=0, \gamma, N=\infty) = \frac{2A}{B(1-\alpha_3, 1/2+\alpha_3)} \times \frac{\Theta(\omega_2(\gamma)-\omega)}{\omega^{2\alpha_3}(\omega_2(\gamma)^2-\omega^2)^{1/2-\alpha_3}}, \tag{2.11}$$

where $B(x,y)$ is the betafunction. In this ansatz the high frequency excitations are cut off at

$$\omega_2(\gamma) = 2\pi\frac{\sin\gamma}{\gamma}. \tag{2.12}$$

The ansatz coincides with the exact result of Katsura et al.[4] in the XX-limit $\gamma = \pi/2$ and leads to the following expression for the ratio (2.6):

$$R_3(\tau,\gamma) = \frac{2}{B(1/2-\alpha_3, 1/2+\alpha_3)} \int_0^1 dx\, x^{-2\alpha_3}(1-x^2)^{\alpha_3-1/2} e^{-x\omega_2\tau} \tag{2.13}$$

The integral (2.13) is represented in figures 1(a),(b) by the dotted curves. For the XX-case $\gamma = \pi/2$, we find good agreement of our determination of $R_3(\tau,\gamma)$ with the exact result (2.13). This agreement supports our hypothesis, that the finite size effects can be parametrized adequately by the ansatz (2.10). For $\gamma = 0.3\pi$, our finite size analysis leads to a scaling curve $R_3(\tau,\gamma)$ which differs significantly from the prediction (2.13). This discrepancy might originate from the sharp cutoff (2.12) in the high frequency excitations.

## 3. The Nonintegrable Infrared Singularity in the Transverse Structure Factors at $p = \pi$ and $T = 0$

Let us assume that the transverse structure factor can be split into two parts. The first one contains the nonintegrable infrared singularity ((1.4) for $j = 1$):

$$S_1(\omega, p=\pi, T=0, \gamma, N\to\infty) = \frac{\Theta(\omega-\omega_1)}{(\omega^2-\omega_1^2)^{\alpha_1}} e^{-\omega\tau_0}\tilde{A}(\omega/\omega_1,\gamma) + S_1'(\omega,\gamma). \tag{3.1}$$

The second one is assumed to be free of such a singularity. We have introduced an exponential cutoff for the high frequency contributions in the first term with a parameter $\tau_0$, which will be fixed below. The factor $\tilde{A}(\omega/\omega_1, \gamma)$ is supposed to describe the approach to the infrared singularity. For $\omega_1 \to 0$, $\tilde{A}(\omega/\omega_1, \gamma)$ is assumed to converge to a nonvanishing value for the residue $\tilde{A}(\infty, \gamma)$ of the infrared singularity. Starting from (3.1) we find for the ratio (2.6) in the limit $\omega_1 \to 0$:

$$R_1(\tau, \gamma, N \to \infty) = \frac{I_1(\omega_1(\tau + \tau_0), \gamma) + \omega_1^{2\alpha_1 - 1} I_1'(\tau, \gamma)}{I_1(\omega_1 \tau_0, \gamma) + \omega_1^{2\alpha_1 - 1} I_1'(0, \gamma)}, \tag{3.2}$$

where:

$$I_1(x, \gamma) = \int_1^\infty dy\, (y^2 - 1)^{-\alpha_1} e^{-xy} \tilde{A}(y, \gamma) \tag{3.3}$$

and

$$I_1'(\tau, \gamma) = \int_0^\infty d\omega\, S_1'(\omega, \gamma) e^{-\omega\tau}. \tag{3.4}$$

In the combined limit:

$$N \to \infty, \quad \tau \to \infty, \quad x = \omega_1(\tau + \tau_0) \quad \text{fixed}, \tag{3.5}$$

we expect the ratio (3.2) to converge to a scaling function $I_1(x, \gamma)/I_1(0, \gamma)$, since $2\alpha_1 - 1 > 0$ for $0 < \gamma \leq \pi/2$. The scaling function (3.3) depends on the parametrization of the first contribution in (3.1). The cutoff parameter $\tau_0$ enters in the finite size corrections to the scaling variable $x$. The scaling curve depends explicitly on $\tilde{A}(y, \gamma)$. The scaling behaviour of the ratio (3.2) in the combined limit (3.5) has an immediate consequence for the 'half width' :

$$\ln(\tau(R_1, \gamma, N) + \tau_0) = -\ln \omega_1(N) + \ln x(R_1, \gamma) + ... \tag{3.6}$$

which diverges for $N \to \infty$. Therefore on finite systems, the signature for the emergence of the nonintegrable infrared singularity is a linear increase of the half width with $\ln N$ and with slope 1. This behaviour should be observable not only for $R_1 = 0.5$ but for any fixed value of $R_1$ between 0 and 1. In figure 4 we have plotted the left hand side of (3.6) with $\tau_0 = 0$ versus $-\ln \omega_1(N)$ for $R_1 = 0.75, 0.5, 0.25$ and $\gamma/\pi = 0.5$. The linear behaviour in $-\ln \omega_1(N)$ is clearly seen and the slopes $0.82, 0.92, 0.96$ are found to be rather close to the expectation, namely 1. The second term in (3.6) ($x(R_1, \gamma)$) is the inverse of the scaling function $I_1(x, \gamma)/I_1(0, \gamma)$. The behaviour of the scaling function for small values of the scaling variable $x = \omega_1(\tau + \tau_0)$:

$$I_1(x \to 0, \gamma) - I_1(0, \gamma) = -\tilde{A}(\infty, \gamma)\Gamma(1 - 2\alpha_1) x^{2\alpha_1 - 1}, \tag{3.7}$$

is governed by the exponent $\alpha_1$ of the infrared singularity.

So far our discussion of the nonintegrable infrared singularity is restricted to the transverse case with anisotropy $\pi/2 \geq \gamma > 0$. In the isotropic case $\gamma = 0, \alpha_1 = \alpha_3 = 0.5$ the integral (3.3) diverges logarihmically for $x \to 0$:

$$I_1(x, \gamma = 0) = -\tilde{A}(\infty, \gamma = 0) \ln x \tag{3.8}$$

From (3.8) we predict that the 'half width' in the isotropic case:

$$\ln[\tau(R_1, \gamma = 0, N) + \tau_0] = (R_1 - 1) \ln \omega_1(N) + C(R_1) + ... \tag{3.9}$$

increases with $\ln N$ but with a slope $(1 - R_1)$ depending on $R_1$. Again this behaviour is visible even on small systems, if we make a proper choice for $\tau_0 = 0.2$, as can be seen from figure 5. The observed slopes have the correct $R_1$-dependence. Their absolute value differs from the expectation by about 10%.

## 4. Dynamical Structure Factors in the Noncritical Regime

Leaving the critical momentum $p = \pi$ the threshold singularities in the dynamical structure factors change in position and strength. The singularity moves according to (1.3) to nonvanishing frequencies $\omega = \omega_1(p, \gamma)$; its strength is reduced to $(\omega - \omega_1)^{-\alpha_j}$. The threshold singularity is integrable now, since $\alpha_j < 1$ for $j = 1, 3$ and the Laplace transforms of the transverse and longitudinal structure factors exist for all nonnegative $\tau$-values, if $0 < p < \pi$. The ratios:

$$R_j(\tau, p, \gamma, N) = \frac{S_j(\tau, p, T = 0, \gamma, N)}{S_j(\tau = 0, p, T = 0, \gamma, N)} \stackrel{N\to\infty}{\longrightarrow} R_j(\tau, p, \gamma) \tag{4.1}$$

converge to a limiting function in $\tau$. Its large $\tau$-behaviour is given by the threshold frequency (1.3) and the strength of the threshold singularity:

$$R_j(\tau, p, \gamma) \stackrel{\tau\to\infty}{\longrightarrow} \exp(-\omega_1 \tau) \tau^{\alpha_j - 1}. \tag{4.2}$$

We have determined the ratios $R_1$ and $R_3$ as function of $\tau$ at fixed noncritical momenta and for various values of $\gamma$. We found extremely small finite size effects. Moreover the ratios (4.1) almost coincide for the different values of $\gamma$ and for the longitudinal and transverse case. We only observe a weak dependence on the momentum $p$. As an example we present in figure 6 the results for the isotropic case ($\gamma = 0$) at fixed momenta $p/\pi = 1/4, 3/4, 1/2$. The first two momenta can be realized for $N = 8, 16$. Here finite size effects cannot be resolved in the plot of the ratio (4.1). The origine of the $\ln \tau$-axis has been shifted by 2 in order to present the results for $p = \pi/2, N = 4, 8, 12, 16$. Again the ratios (4.1) coincide for $N = 8, 12, 16$, whereas the result for $N = 4$ is found left to them.

Significant finite size effects only appear for large values of $\tau$ and are visible in the quantities $\rho_j(t_j, \gamma, N)$ which are related to the ratios (4.1) via:

$$R_j(\tau, \gamma, N) = [1 + \rho_j(t_j, \gamma, N)]^{-1}, \tag{4.3}$$

where

$$t_j = \tau^{1-\alpha_j} \exp(\omega_1 \tau). \tag{4.4}$$

According to (4.2), $\rho_j(t_j, \gamma, N = \infty)$ should be linear in $t_j$ for $t_j \to \infty$. In figure 7 we show this quantity for the isotropic case at $p = \pi/2$ and $N = 4, 8, 12, 16$. The linear behaviour in $t$ becomes more and more apparent with increasing system size. We can compare our results for the longitudinal case with the prediction:

$$R_3(\tau, p, T = 0, \gamma) = \frac{I_3(\tau, p, \gamma)}{I_3(\tau = 0, p, \gamma)}, \tag{4.5}$$

$$I_3(\tau, p, \gamma) = \int_{\omega_1}^{\omega_2} d\omega (\omega^2 - \omega_1^2)^{-\alpha_3} (\omega_2^2 - \omega^2)^{-1/2+\alpha_3} e^{-\omega\tau} \tag{4.6}$$

which follows from the ansatz of Müller et al.[10]. In (4.6) the lower and upper integration bounds are given by (1.3) and:

$$\omega_2(p, \gamma) = \frac{2\pi}{\gamma} \sin\gamma \sin\frac{p}{2}, \tag{4.7}$$

respectively. The prediction is found between the finite system results for $N = 12$ and $N = 16$ (cf. the dotted curve in figure 7). Better agreement with the expected behaviour in the thermodynamical limit can be achieved for example by lowering the upper integration bound $\omega_2$ in equation (4.6). This, however, would violate the sumrules of Ref. [13]. To our knowledge there does not yet exist a generalization of the ansatz of Müller *et al.* [10], which respects the sumrules in Ref. [13].

In figure 8 we show the ratio $R$ as fuction of the momentum $p$ $(0 < p < \pi)$ and at fixed: $\tau = 0.5, \quad \gamma/\pi = 0.0 \quad N = 4, 6, ..., 16$. All these data points nicely follow one curve in the 'scaling'-variable $p$. Again this means that the thermodynamical limit is seen already on small systems.

## 5. Conclusions

In this paper we started a first attempt to extract the dynamical structure factors of the XXZ-model from a complete diagonalization of the Hamiltonian on finite rings with $N = 4, 6, ...16$ sites. We studied the normalized ratios (2.6) of the dynamical structure factors as function of the euclidean (imaginary) time $\tau$ and found the following features:

(i) Away from the critical momentum $p = \pi$ finite size effects are small except for the large $\tau$-limit, where we find a clean signal for the threshold singularity (1.4).

(ii) At the critical momentum $p = \pi$ finite size effects are still small in the longitudinal case at $\gamma = \pi/2$ but increase for decreasing values of $\gamma$. These finite size effects can be described adequately by the ansatz (2.10). The resulting thermodynamical limit (2.8) for the ratio (2.6) is in good agreement with the exact result of Katsura et al.[4] at $\gamma = \pi/2$ but deviates from the prediction (2.13) of Müller et al. [10] for smaller values of $\gamma$.

(iii) At the critical momentum $p = \pi$ finite size effects are large in the transverse case and increase rapidly with increasing values of $\gamma$. It was demonstrated in section 3 that this behaviour signals the emergence of the (nonintegrable) infrared singularity in the tranverse structure factor.

Therefore we can conclude that the euclidean time representation is particularly suited for the study of finite size effects, which allows for a crude estimate of the thermodynamical limit. This estimate is already precise enough to check the gross features of a model ansatz [10] for the dynamical correlation functions in the spectral ($\omega$)-representation. To resolve the fine structure, however, - i.e. the detailed form and cutoff of the nonsingular contributions- one has to know the euclidean time dependence very precisely. In other words: The reconstruction of the spectral ($\omega$)-representation from the euclidean time ($\tau$)-representation demands for an anlytic continuation and tiny errors in the $\tau$- might produce large errors in the $\omega$-representation. Of course this problem would not occur if finite size effects could be analyzed systematically in the spectral representation. The continued fraction method used in Reference [14] might open this possibility.

Our exact results on small systems might be useful as a test for those, who plan to develop approximative calculations on larger systems. For this purpose we present in appendix A our results for the first nine excitation energies with the corresponding transition probabilities at fixed $p/\pi = 1/4, 1/2, 3/4, 1; N = 16; \gamma = 0$.

## Appendix A. Excitation Energies and Transition Probabilities for $N = 16$

In the following table we present the excitation energies $\omega_n = E_n - E_0$ and the transition probabilities $t_n = |\langle n|s_3(0)|0\rangle|^2$ for the isotropic case at fixed momenta $p/\pi = 1/4, 1/2, 3/4, 1$:

| $S(\tau=0, p=\pi/4) = 2.98276632317824 \cdot 10^{-1}$ | | $S(\tau=0, p=\pi/2) = 0.679437576126672$ | |
|---|---|---|---|
| $\omega_n(\pi/4)$ | $t_n(\pi/4)$ | $\omega_n(\pi/2)$ | $t_n(\pi/2)$ |
| 2.30261899538436 | $2.94237604448345 \cdot 10^{-1}$ | 3.38066138588931 | $5.84557616588252 \cdot 10^{-1}$ |
| 4.29305867857873 | $3.26317936539563 \cdot 10^{-4}$ | 4.19713536357146 | $8.75335130248142 \cdot 10^{-2}$ |
| 4.55019377043269 | $2.55896169667915 \cdot 10^{-3}$ | 4.59757074255291 | $4.16484504934824 \cdot 10^{-3}$ |
| 5.00373864011045 | $2.16897422554017 \cdot 10^{-4}$ | 4.73871528573011 | $1.00961950227532 \cdot 10^{-3}$ |
| 5.56892568932023 | $8.32876542259278 \cdot 10^{-5}$ | 4.90280915101872 | $6.60897218681834 \cdot 10^{-5}$ |
| 5.78139338722018 | $1.30702193321460 \cdot 10^{-4}$ | 5.35132697794261 | $1.17091017648893 \cdot 10^{-4}$ |
| 5.84500569454930 | $4.97880013858312 \cdot 10^{-4}$ | 5.89030536873391 | $2.04831697796575 \cdot 10^{-4}$ |
| 6.29013618714556 | $2.13454771246274 \cdot 10^{-5}$ | 5.92901985838870 | $1.27434335470092 \cdot 10^{-3}$ |
| 6.33212143869286 | $1.26096012286078 \cdot 10^{-5}$ | 6.47575389511423 | $5.63517350683256 \cdot 10^{-5}$ |

| $S(\tau=0, p=3\pi/4) = 1.32305343430234$ | | $S(\tau=0, p=\pi) = 4.29230350827985$ | |
|---|---|---|---|
| $\omega_n(3\pi/4)$ | $t_n(3\pi/4)$ | $\omega_n(\pi)$ | $t_n(\pi)$ |
| 2.63813043456811 | 1.02102198318191 | 0.54037936450057 | 3.43961688789893 |
| 3.41153204282532 | $1.02545237335683 \cdot 10^{-3}$ | 2.79206117219889 | $6.05607890324550 \cdot 10^{-1}$ |
| 4.33065574211427 | $2.45752087828875 \cdot 10^{-1}$ | 4.66859660533213 | $2.00416101651547 \cdot 10^{-1}$ |
| 5.03583236201723 | $1.18879754929133 \cdot 10^{-3}$ | 5.47594709145050 | $7.02305448815108 \cdot 10^{-4}$ |
| 5.44067886083706 | $5.87109321145863 \cdot 10^{-4}$ | 5.90701658134804 | $4.48503627312539 \cdot 10^{-2}$ |
| 5.46270297857887 | $5.22628990055292 \cdot 10^{-2}$ | 5.99408178622140 | $4.45448599279042 \cdot 10^{-6}$ |
| 5.98040322135934 | $1.06884234982829 \cdot 10^{-4}$ | 6.57325345602260 | $6.66105581887054 \cdot 10^{-4}$ |
| 6.44585572118417 | $3.08052603394593 \cdot 10^{-5}$ | 6.80283246183318 | $2.87699128786244 \cdot 10^{-4}$ |
| 6.64303864126450 | $7.25411318058785 \cdot 10^{-4}$ | 7.14462784940782 | $1.98076177494986 \cdot 10^{-6}$ |

[1] Fabricius K, Löw U, Mütter K-H and Ueberholz P, Phys. Rev. B**44**, 7476, (1991)
Fabricius K, Löw U, and Mütter K-H, Z. Phys. B**91**, 51, (1993).

[2] Fabricius K, Löw U, and Mütter K-H , *'Complete Solution of the XXZ- Model on Finite Rings. Finite Temperature Structure Factors'*, to be published in Phys. Rev. B

[3] Niemeijer T, Physica **36**, 377, (1967)

[4] Katsura S, Horiguchi T, and Suzuki M, Physica **46**, 67, (1967)

[5] Mazur P, and Siskens T J, Physica **69**, 259, (1973); 71, 560, (1974)

[6] Barouch E, McCoy B M, and Abraham D B, Phys. Rev. A**4**, 2331,(1971)
McCoy B M, Perk J H H, and Schrock R E, Nucl. Phys. B**220** **[FS8]**, 35 (1983)
Müller G and Schrock R E, Phys. Rev. B**29**, 228 (1984)
Stolze J, Nöppert A, and Müller G: *'Gaussian, exponential, and power-law decay of time-dependent correlation functions in quantum spin chains'* january 1995.

[7] Lieb E, Schultz T, and Mattis D, Ann. Phys. (N.Y.), **16**, 407, (1961)

[8] Cloizeaux des J, and Pearson J J, Phys. Rev. **128**, 2131, (1962)

[9] Luther A and Peschel I, Phys. Rev. B**12**, 3908, (1975)
Fogedby H C, J. Phys. C**11**, 4767, (1978)

[10] Müller G, Thomas H, Puga M W, and Beck H, J. Phys. C**14**, 3399, (1981)
Müller G, Thomas H, Beck H, and Bonner J C, Phys. Rev. B**24**, 1429, (1981)

[11] Cowley R A, Tennant D A, Perring T G, Nagler S E, and Tsvelik A M: 'Spin Dynamics of a Heisenberg $s = 1/2$ Antiferromagnetic Chain'

[12] Karbach M , Mütter K-H and Schmidt M, Phys. Rev. B**50**, 9281, (1994)

[13] Müller G, Phys. Rev. B**26**, 1311 (1982)

[14] Viswanath V S, Zhang S, and Müller G, Phys. Rev. B**49**, 9702 (1994)
Viswanath V S and Müller G: *'The Recursion Method-Application to Many-Body Dynamics', Lecture Notes in Physics*, m23, Springer 1994

## Figure Captions

**Figure 1** The ratios (2.6) of the dynamical structure factors at $p = \pi$ versus the logarithm of the euclidean time $\tau$: a) longitudinal $\gamma = \pi/2$; b) longitudinal $\gamma = 0.3\pi$. The dashed curves represent the estimate of the thermodynamical limit according to (2.10). The dotted curves are the integrals (2.13)

**Figure 2** Same as figure 1 for the isotropic case $\gamma = 0$.

**Figure 3** Same as figure 1 for a) transverse $\gamma = 1/3\pi$; b) transverse $\gamma = 1/2\pi$.

**Figure 4** The finite size behaviour of $\ln \tau$ along the horizontal lines $R_1 = 1/4, 1/2, 3/4$ shown in figure 3(b)

**Figure 5** The finite size behaviour of $\ln(\tau + \tau_0), \tau_0 = 0.2$ along the horizontal lines $R = 1/4, 1/2, 3/4$ shown in figure 2.

**Figure 6** The ratios (4.1) of the dynamcal structure factors at noncritical momenta versus $\ln \tau$: a) isotropic $p/\pi = 1/4, 3/4, N = 8, 16$; b) isotropic $p/\pi = 1/2, N = 4, 8, 12, 16$

**Figure 7** The quantity $\rho(t, p = \pi/2, \gamma = 0, N)$ - defined in (4.3) - versus $t$ - defined in (4.4). The dashed curve is the prediction (4.5,6)

**Figure 8** The ratio (4.1) of the dynamical structure factors at fixed euclidean time $\tau = 0.5$ versus momentum $p$ and $N = 4, 6, ..., 16$

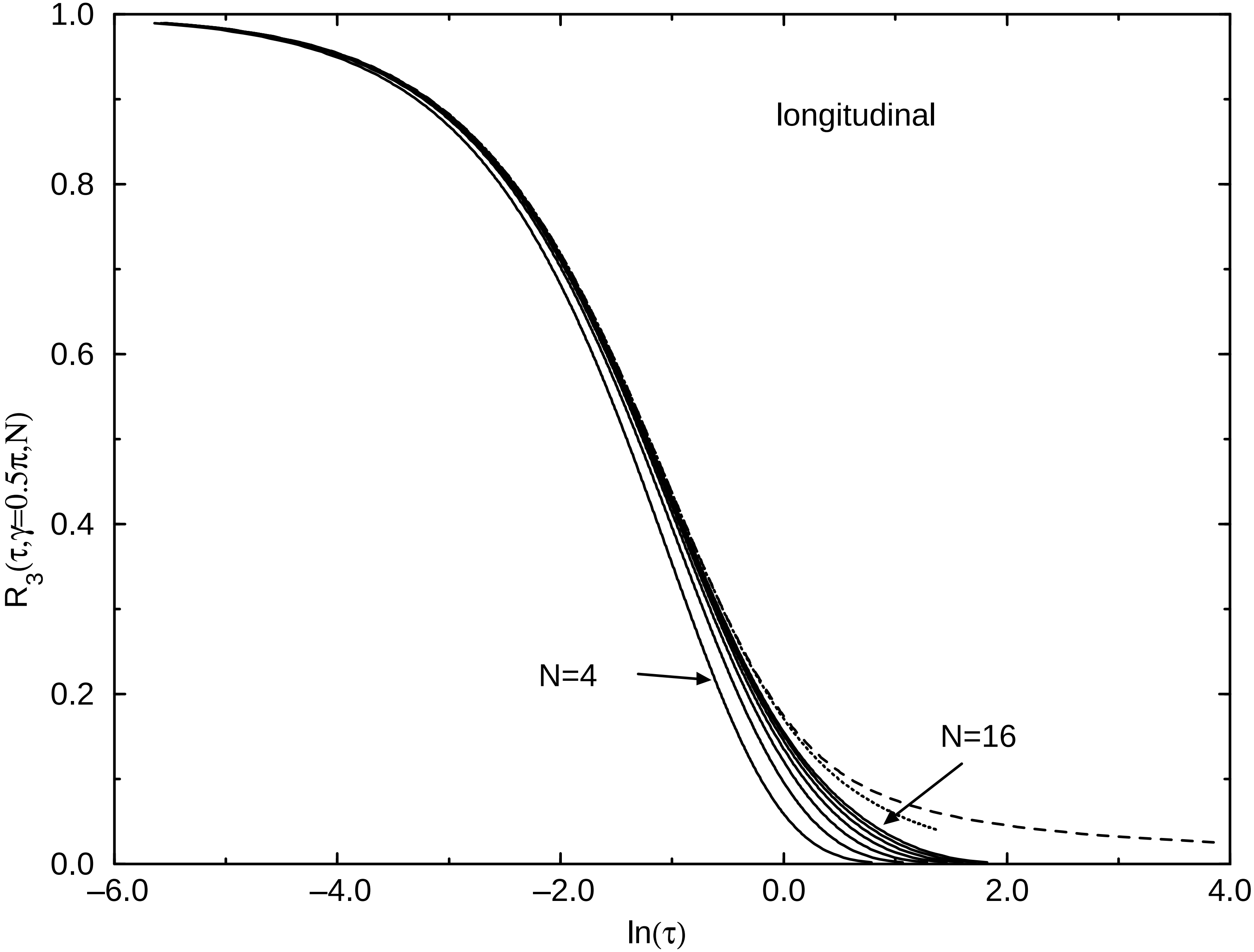

*Fig.1a*

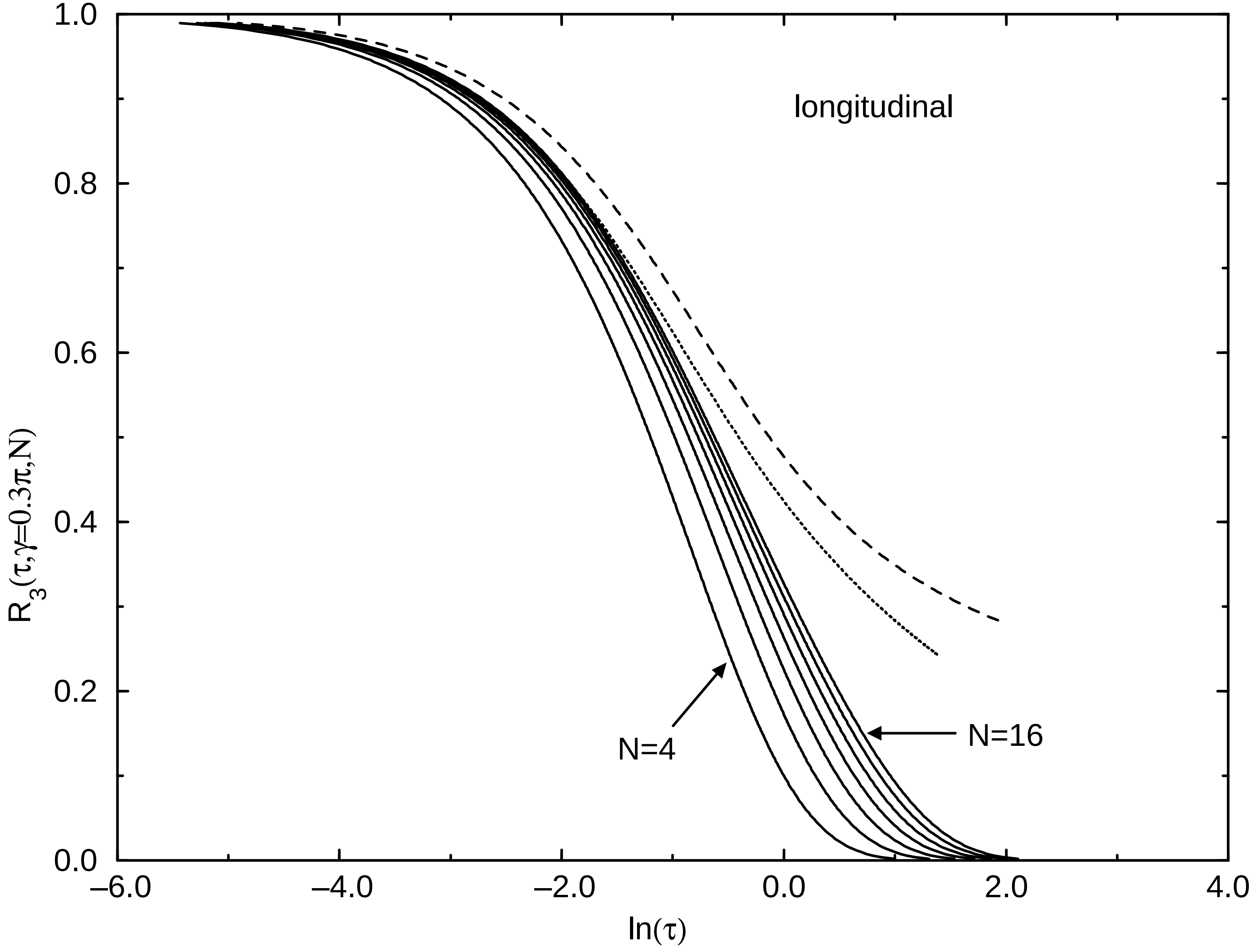

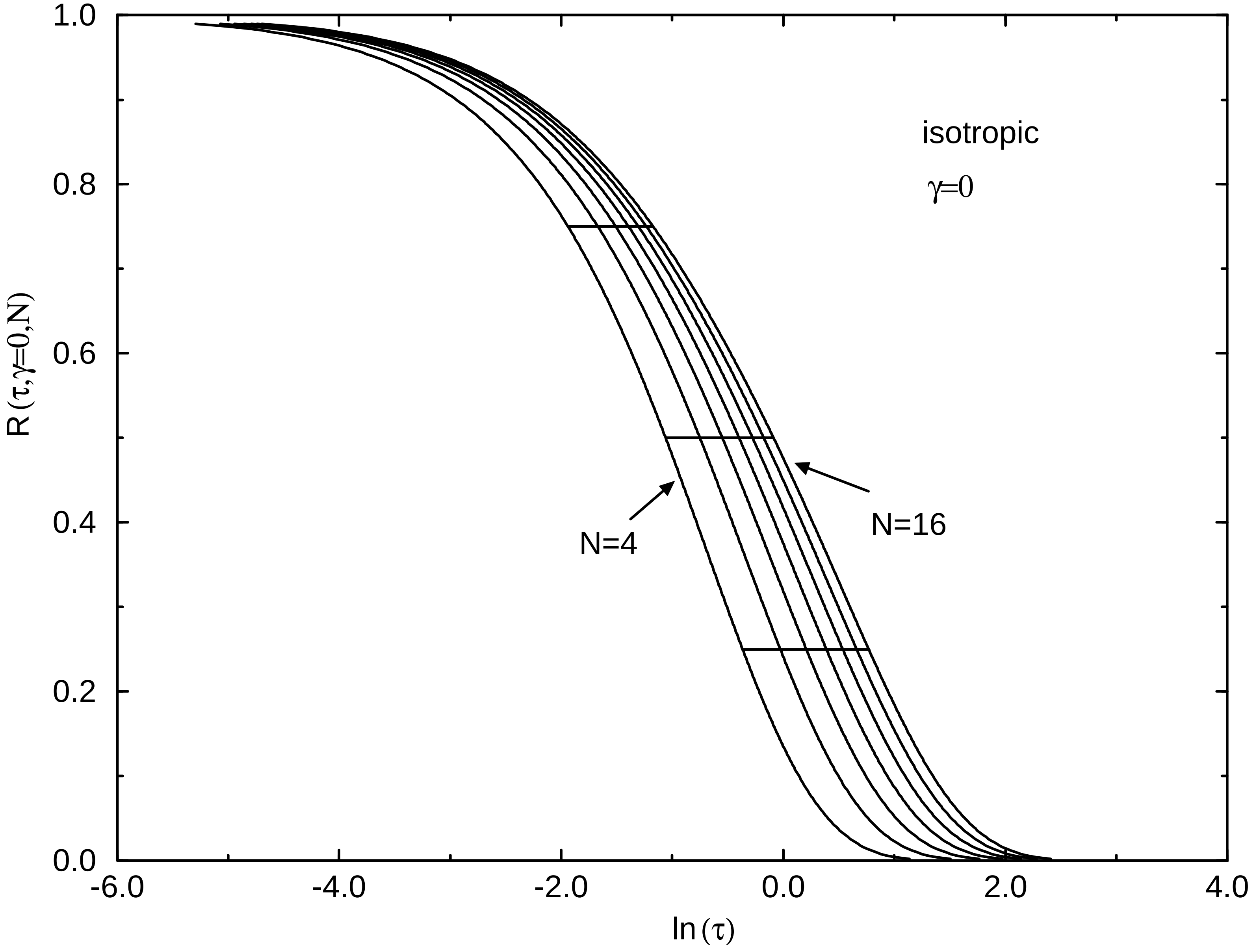

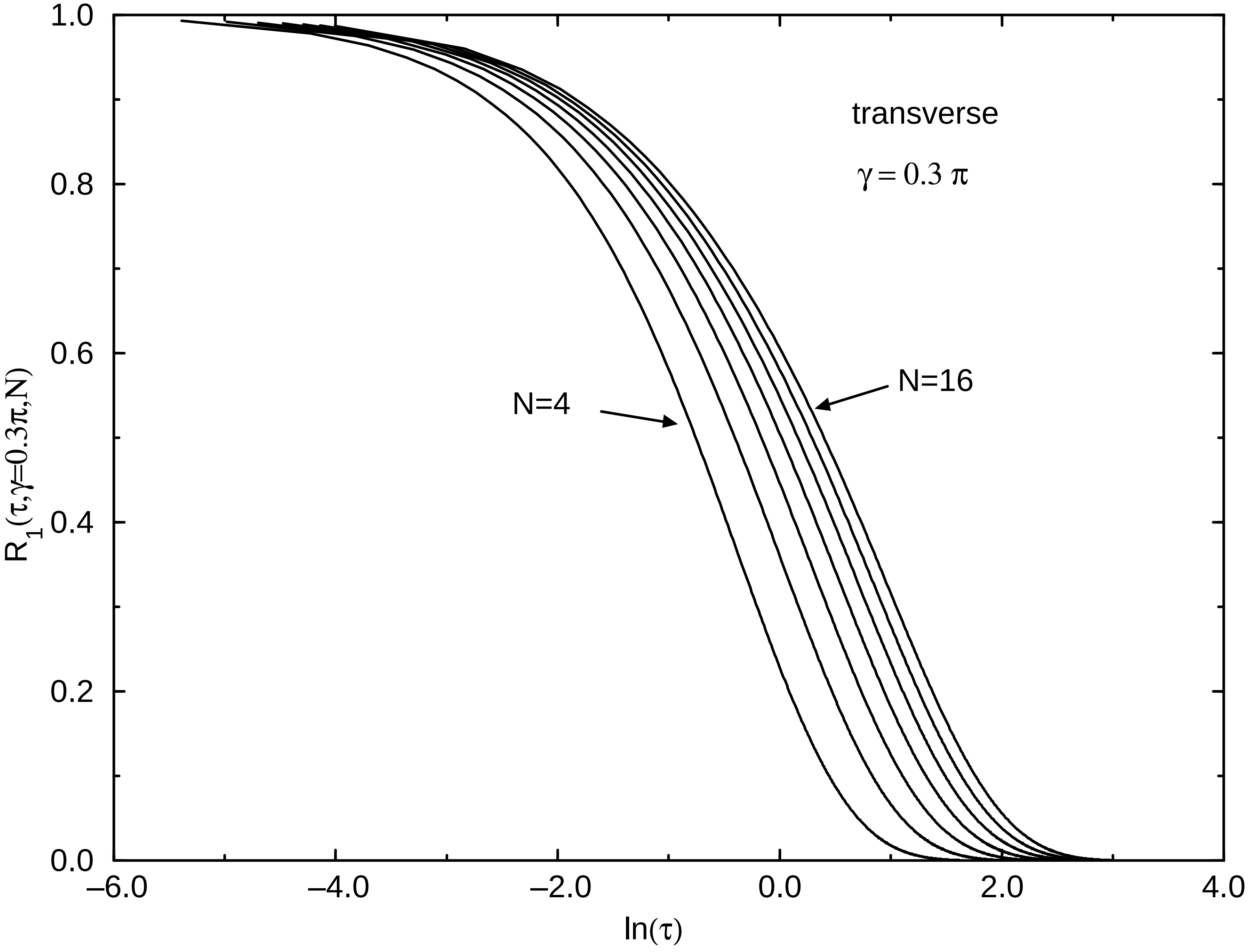

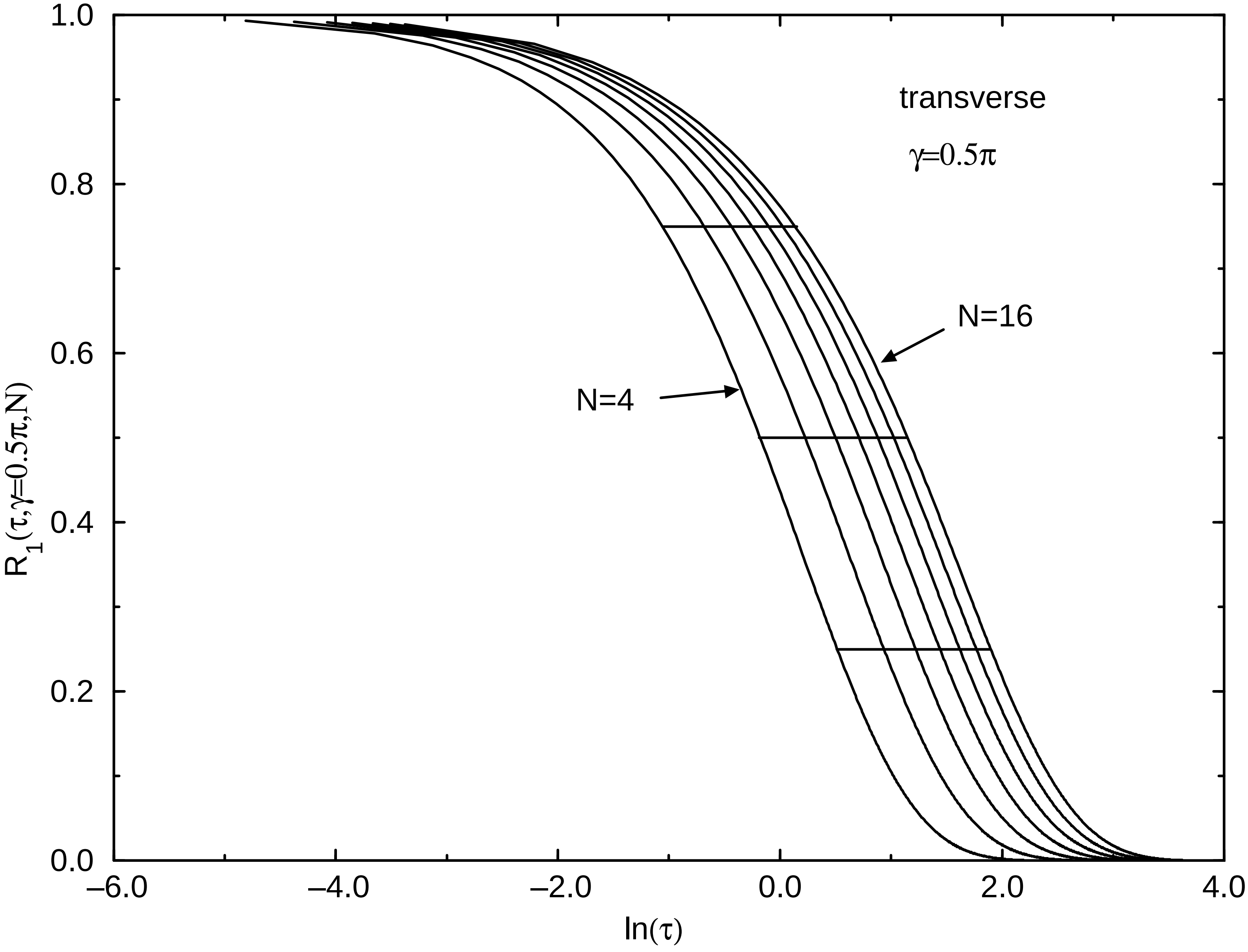

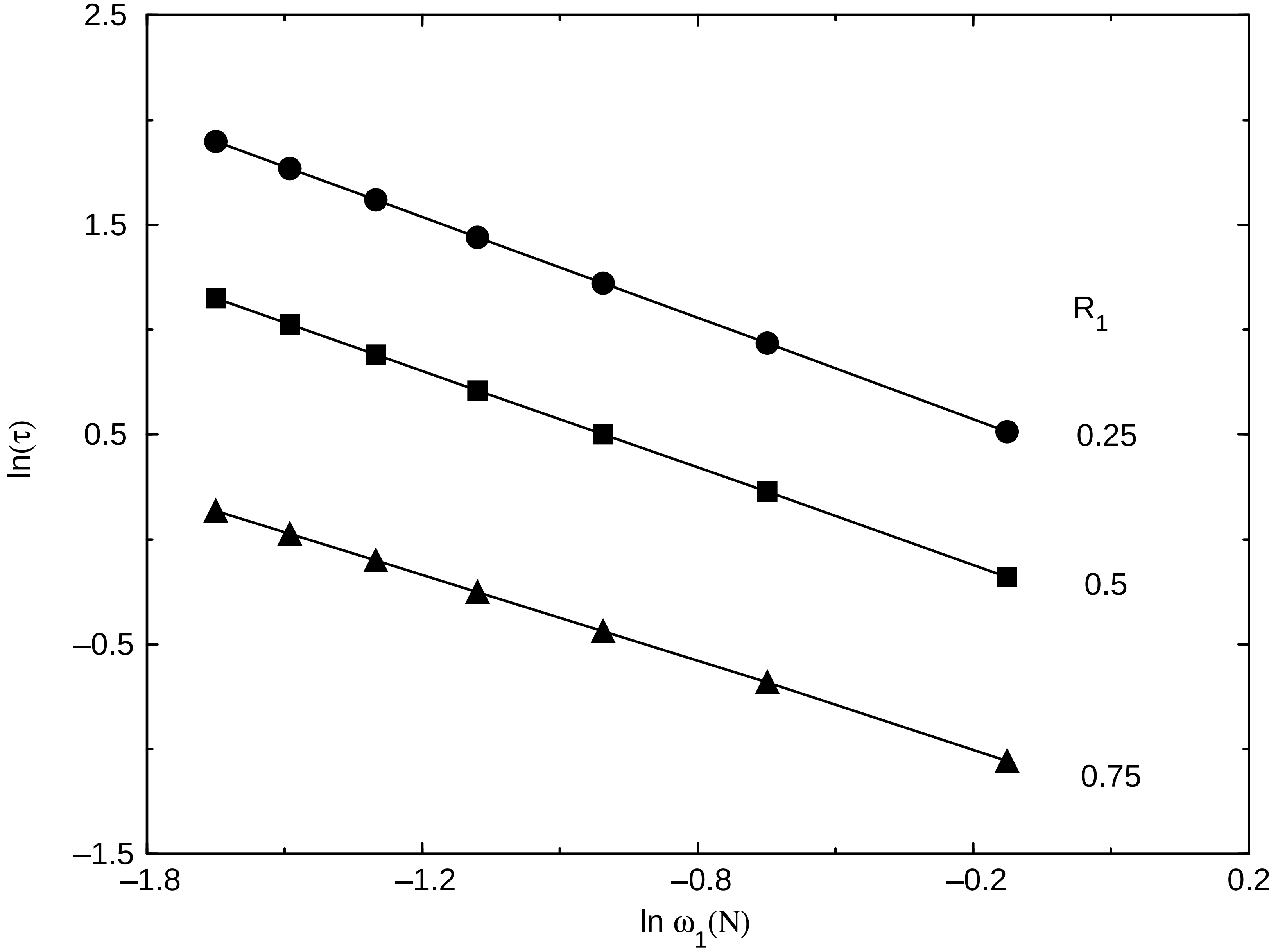

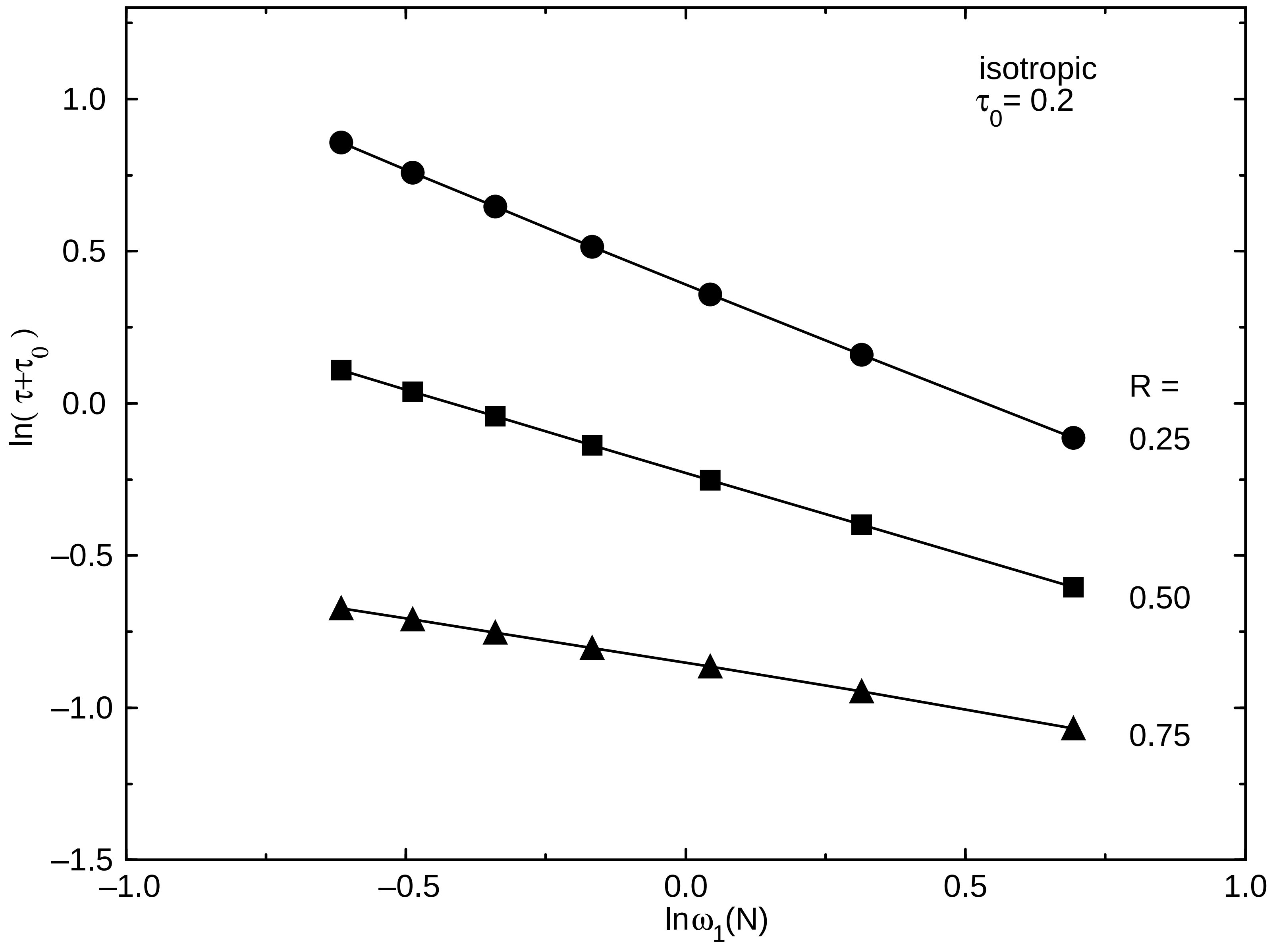

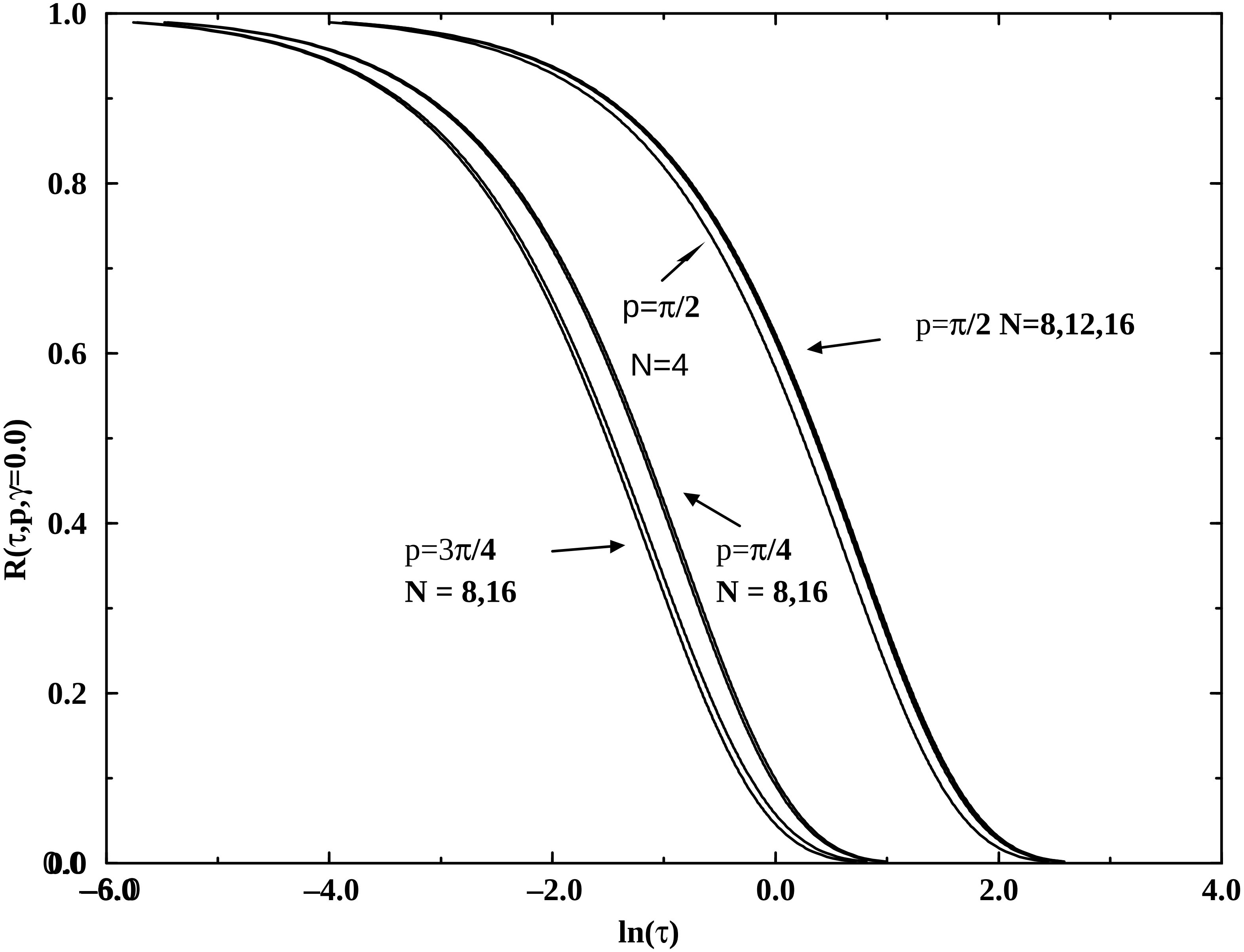

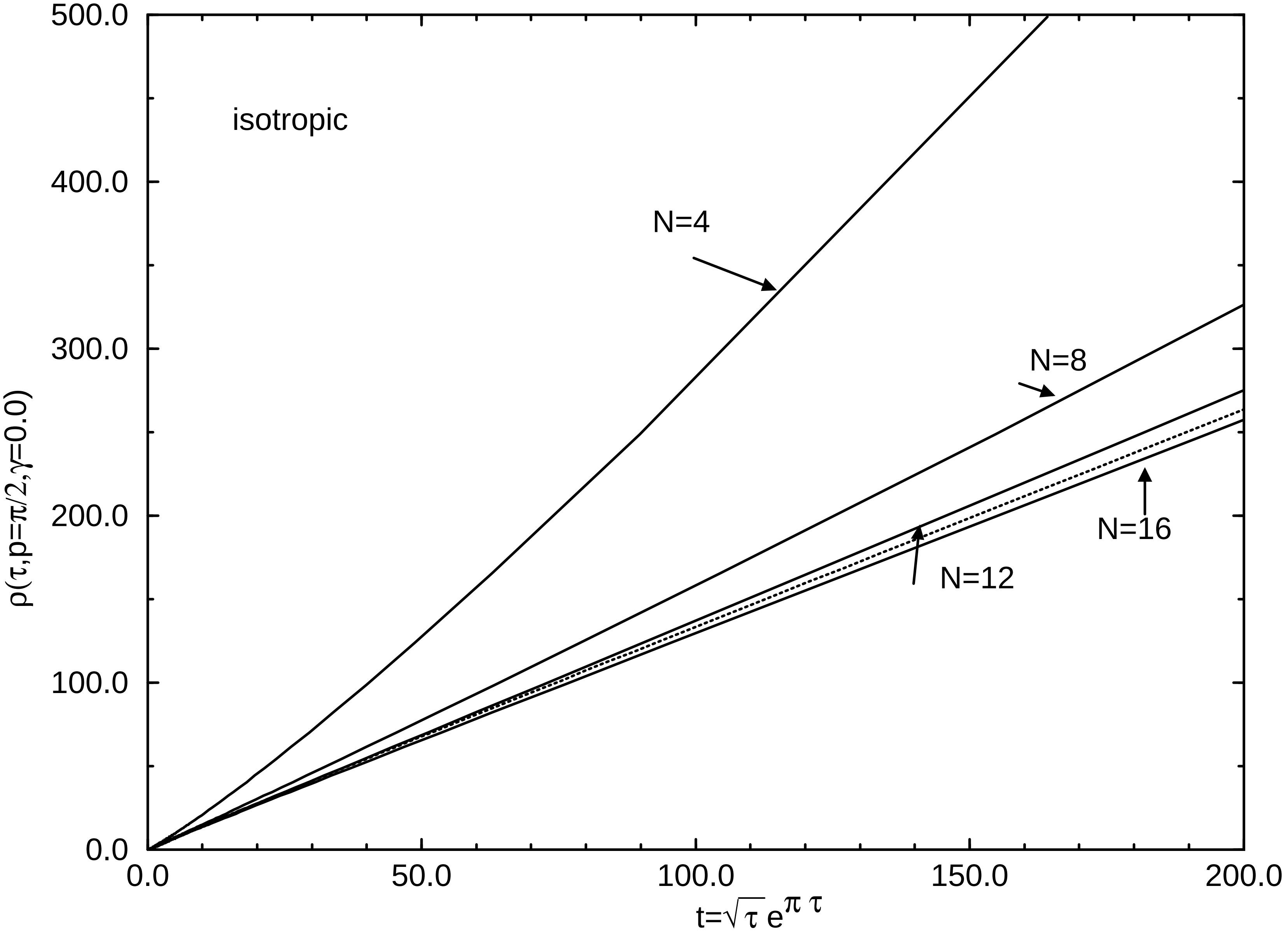

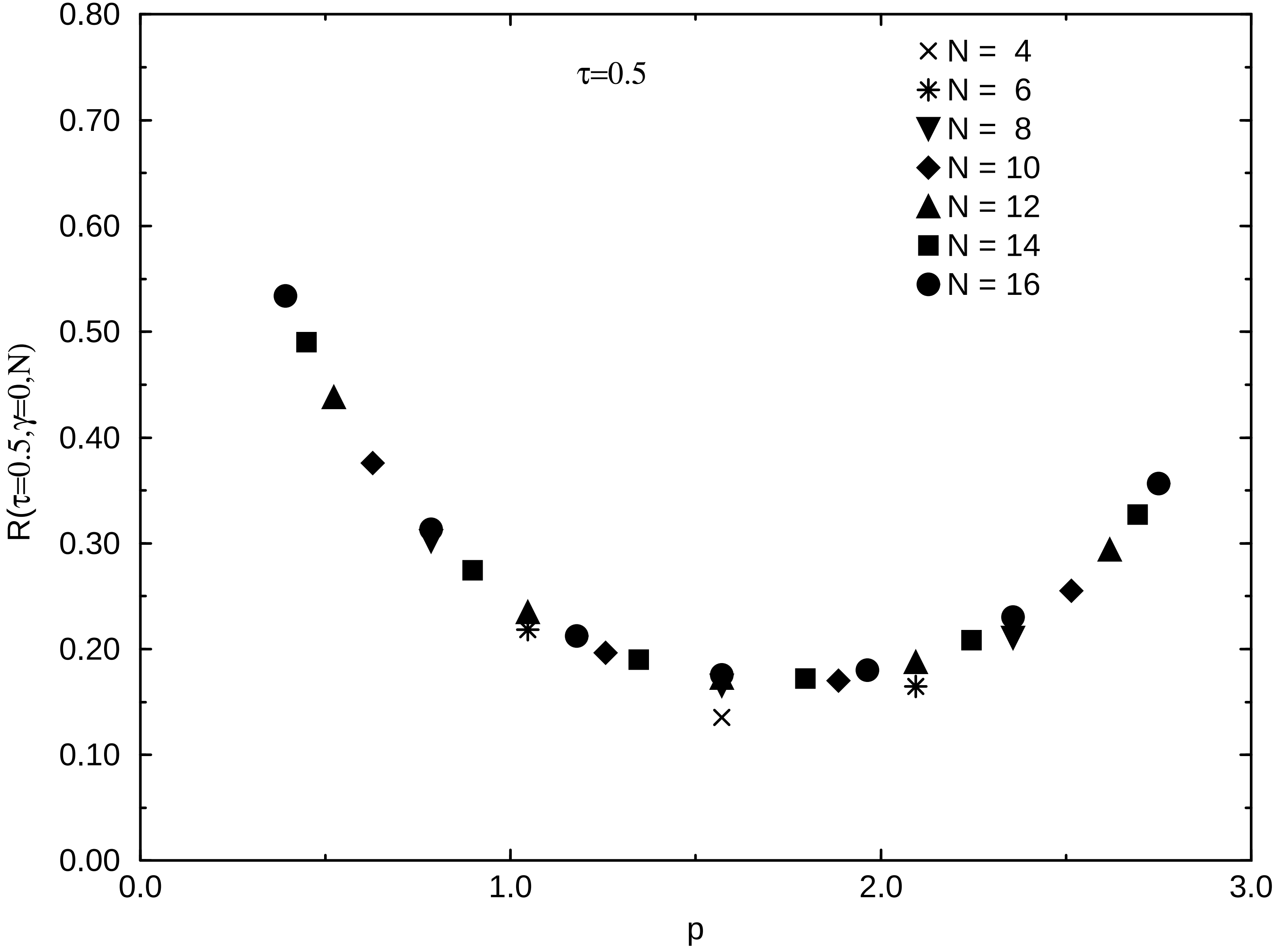